\documentclass[conference]{IEEEtran}
\IEEEoverridecommandlockouts
\usepackage{cite}
\usepackage{float}
\usepackage{amsmath,amssymb,amsfonts}
\usepackage{graphicx, enumitem}
\usepackage{textcomp}
\usepackage[ruled,vlined,linesnumbered]{algorithm2e}
\usepackage[table,dvipsnames]{xcolor}
\usepackage{tikz}
\usepackage{bm}
\usepackage{color,soul}
\usepackage{makecell}
\usepackage{booktabs}
\usepackage{multirow}
\usepackage{url}
\usepackage{subfigure}
\usepackage{tikz}
\usepackage[none]{hyphenat}
\SetKwIF{If}{ElseIf}{Else}{if}{}{else if}{else}{end if}%
 
\setlist[itemize]{leftmargin=4mm}

\newcolumntype{L}[1]{>{\raggedright\let\newline\\\arraybackslash\hspace{0pt}}m{#1}}
\newcolumntype{C}[1]{>{\centering\let\newline\\\arraybackslash\hspace{0pt}}m{#1}}
\newcolumntype{R}[1]{>{\raggedleft\let\newline\\\arraybackslash\hspace{0pt}}m{#1}} 

\def\BibTeX{{\rm B\kern-.05em{\sc i\kern-.025em b}\kern-.08em
    T\kern-.1667em\lower.7ex\hbox{E}\kern-.125emX}}

\expandafter\def\expandafter\normalsize\expandafter{%
    \normalsize%
    \setlength\abovedisplayskip{4pt}%
    \setlength\belowdisplayskip{4pt}%
}

\begin{document}

\title{Exploring Utility in a Real-World Warehouse Optimization Problem: Formulation Based on Quantum Annealers and Preliminary Results\\
	{}
	\thanks{This work was supported by the Basque Government through HAZITEK program (Q4\_Real project, ZE-2022/00033) and through Plan complementario comunicación cuántica (EXP. 2022/01341) (A/20220551).}
}

\author{
	\IEEEauthorblockN{Eneko Osaba\IEEEauthorrefmark{2}\IEEEauthorrefmark{1}, 
        Esther Villar-Rodriguez\IEEEauthorrefmark{2}, and
	    Antón Asla\IEEEauthorrefmark{3}}
	\IEEEauthorblockA{\IEEEauthorrefmark{2}TECNALIA, Basque Research and Technology Alliance (BRTA), 48160 Derio, Bizkaia, Spain}
    \IEEEauthorblockA{\IEEEauthorrefmark{3}Serikat - Consultoría y Servicios Tecnológicos, 48009 Bilbao, Spain.}
	\IEEEauthorblockA{\IEEEauthorrefmark{1}Corresponding author. Email: eneko.osaba@tecnalia.com}}
\maketitle

\IEEEpubidadjcol	

\begin{abstract}
In the current NISQ-era, one of the major challenges faced by researchers and practitioners lies in figuring out how to combine quantum and classical computing in the most efficient and innovative way. In this paper, we present a mechanism coined as \textit{Quantum Initialization for Warehouse Optimization Problem} that resorts to D-Wave's Quantum Annealer. The module has been specifically designed to be embedded into already-existing classical software dedicated to the optimization of a real-world industrial problem. We preliminary tested the implemented mechanism through a two-phase experimentation against the classical version of the software.
\end{abstract}

\section{Introduction}\label{sec:intro}

Being immersed in the NISQ-era, quantum-classical hybrid systems are destined to play a major role in the adoption of quantum computing to deal with industrial problems \cite{osaba2024hybrid}. Within this framework, the challenge lies in figuring out how to combine classical and quantum computing in a way that produces a tandem that ultimately performs better than strictly classical conceptualizations.

This study focuses on a real-world industrial problem named Warehouse Optimization Problem (\texttt{WOP}). This problem describes a warehouse composed of a set of $L$ locations and a number of $I$ items to store. On the one hand, each location $l \in L$ is defined by its capacity $c_l$ (measure in $m^2$), its type (floor or shelf), the time required to place an item on it, and a constant additional time for each height level. On the other hand, each item $i \in I$ is characterized by its type ID and the area $a_{i,l}$ it occupies, which is an integer constant value except for prohibited placements (typically in shelf-type locations) where $a_{i,l}=\infty$. In addition, items of the same type can be stacked up to a maximum height of $h_i$ (where $h_i$=1 if item $i$ cannot be stacked). The main purpose of WOP is to assign a storage location $l$ and a specific height (if stacked) to each item $i$, where the objective function is both to minimize the storage time ($o_1$) and the total area occupied ($o_2$). 

\begin{figure}[b]
    \centering
    \includegraphics[width=1.0\linewidth]{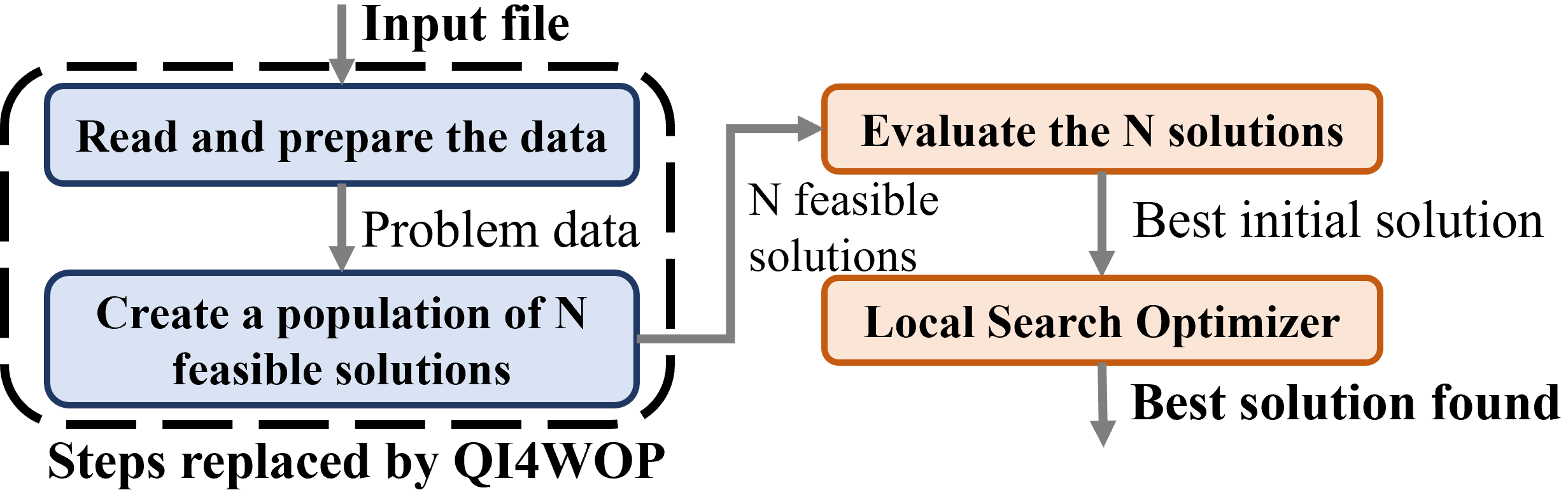}
    \hspace*{-1cm}
    \caption{Workflow of the full-classical PoC.}
    \label{fig:PoC}
\end{figure}

This problem has been identified by a Spanish company called Ertransit, which already counts with a classical Proof-of-Concept (PoC) algorithm to solve \texttt{WOP}. This PoC, which workflow is depicted in Figure \ref{fig:PoC}, is composed of two modules: i) \textit{initialization module}, devoted to creating a set of random feasible solutions to \texttt{WOP}; and ii) \textit{local-search module}, which selects the best solution generated in the previous step and optimizes it using a local search optimization technique.

The tests conducted have shown that, despite promising results, this PoC has room for improvement. Taking this opportunity as our main motivation, this paper presents a quantum-classical mechanism coined as \textit{Quantum Initialization for Warehouse Optimization Problem} (\texttt{QI4WOP}), which resorts to the \textit{Leap Constrained Quadratic Model (CQM) Hybrid Solver} (\texttt{LeapCQMHybrid}) of D-Wave \cite{leapCQM}.




\section{Description of \texttt{QI4WOP} module}\label{sec:method}

\begin{figure}[t]
    \centering
    \includegraphics[width=0.95\linewidth]{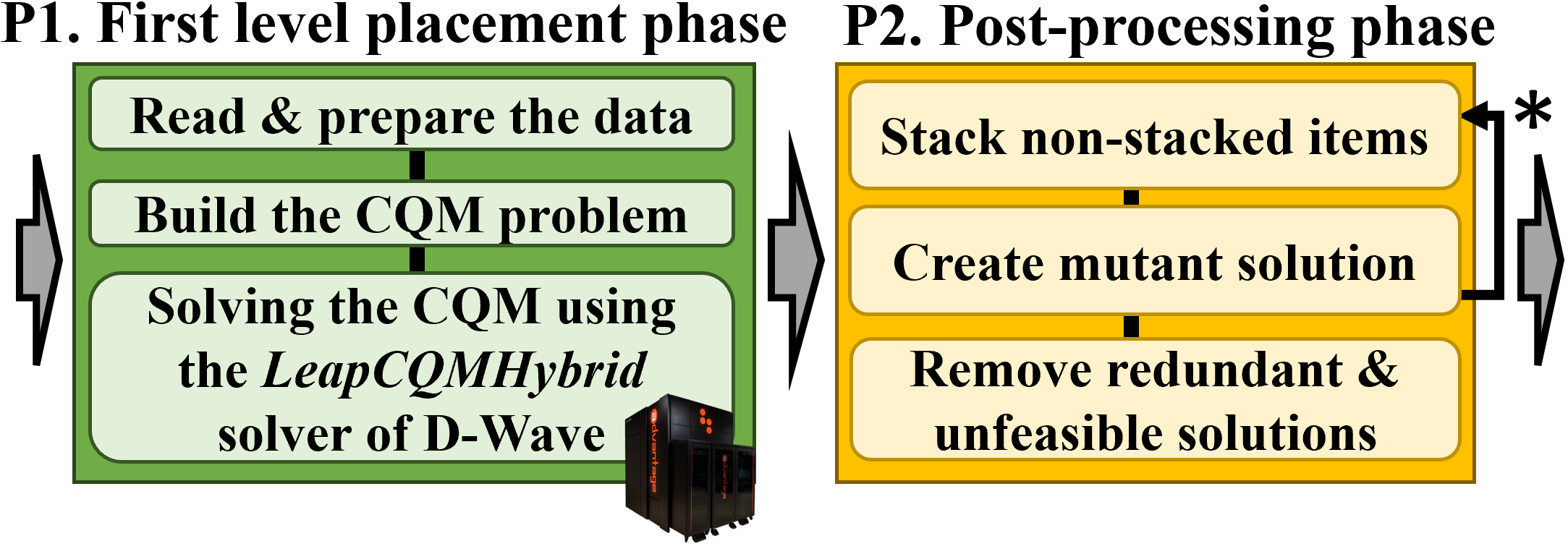}
    \vspace{-0.2cm}
    \caption{Workflow of \texttt{QI4WOP}. \textbf{Input}: input problem. \textbf{Output}: $M$ \texttt{WOP} feasible solutions. * the first two steps of \textbf{P2} are iteratively executed for each of the $N$ partial solutions generated in \textbf{P1}.}
    \label{fig:QI4WOP}
\end{figure}

The tests on the classical PoC have demonstrated that the \textit{initialization module} consumes a lot of computational resources. In response to that, the \texttt{QI4WOP} module presented in this paper substitutes the current \textit{initialization module}. For this reason, \texttt{QI4WOP} is not focused on solving the whole \texttt{WOP} but on finding a large number of feasible solutions to the problem in a reduced computational time. Having said that, \texttt{QI4WOP} module is divided into two phases: \textit{i)} \textit{First level placement phase} and \textit{ii)} \textit{Post-processing phase}. Figure~\ref{fig:QI4WOP} represents the workflow of the \texttt{QI4WOP} module.

 
\textbf{P1. First level placement phase}: This first step is in charge of solving the here called \texttt{sub-WOP} subproblem, whose objective is to store as many items as possible at the ground level of the available set $L$ of locations. This is the step in which \texttt{LeapCQMHybrid} is used. In a nutshell, \texttt{LeapCQMHybrid} deals with problems built as CQM, which is a mathematical model defined by integer, real, and binary variables. CQM admits linear and quadratic restrictions, allowing the introduction of equalities and inequalities and employing quadratic objective functions. 

Regarding the codification, a solution to \texttt{sub-WOP} is represented as a set $X=\{X_1,\dots,X_I\}$ of lists, in which each $X_i$ is associated with a single item $i$. Thus, $X_i=\{x_{i1},\dots,x_{iL}\}$, where $x_{il}$ is 1 if the item $i$ is stored in location $l$, and 0 otherwise. 

The main objective $o$ of \texttt{sub-WOP} is to maximize the amount of items stored, which can be formulated as
\begin{equation}\label{eq:objective}
o = \min -\textstyle \sum_{l=1}^{L}x_{il}\quad\forall i\in \{1,\dots,I\}.
\end{equation}%
Note that the objective is represented as a minimization due to CQM requirements. Furthermore, this objective is subject to four restrictions. The first constraint establishes that an item $i$ must be stored in, at most, one location $l$:
\begin{equation}\label{eq:storing_consistency}
\textstyle \sum_{l=1}^{L}x_{il}\leq1\quad\forall i\in \{1,\dots,I\}.
\end{equation}%
The second restriction assures that items located in one specific location $l$ do not exceed its maximum available area:
\begin{equation}\label{eq:area}
\textstyle \sum_{i=1}^{I}x_{il}*a_{il}\leq c_l\quad\forall l\in \{1,\dots,L\}.
\end{equation}%
The third restriction guarantees that only enabled items can be placed on shelves:
\begin{equation}\label{eq:shelves_consistency}
\textstyle \sum_{l=1}^{L}x_{il}*a_{il}\geq0\quad\forall i\in \{1,\dots,I\}.
\end{equation}%
Last constraint imposes that all non-stackable items are stored.
\begin{equation}\label{eq:non-stackable}
(\textstyle \sum_{i=1}^{I}x_{il})-p_{i}\geq0\quad\forall i\in \{1,\dots,I\}.
\end{equation}%
where $p_i$ is 0 if item $i$ can be stacked and 1 otherwise.

\textbf{P2. Post-processing phase}: when solving \texttt{sub-WOP}, \texttt{LeapCQMHybrid} creates a population of \emph{partial} solutions. Each of them undergoes two classical post-processing steps:

\begin{itemize}
    \item \textbf{Stack non-stacked items}: This step is devoted to completing the \emph{partial} solution by stacking all items that have been left unstacked after \textbf{P1}. First, the number of stacks that can be built for each item type is calculated. Then, each unstacked item is assigned to one stack, respecting the maximum height $h_i$ defined for its corresponding item type. Thanks to this step, the \emph{partial} solution is transformed into a complete \texttt{WOP} solution.
    \item \textbf{Create mutant}: The motivation for this step is to add diversity to the population of solutions. To do this, a new \emph{mutant} solution is created from the previously generated \texttt{WOP} solution, where all items placed at ground level are stacked in a feasible pile with a 50\% probability. Solutions created in this step are added to the population.
\end{itemize}

Lastly, \textbf{P2} ends by eliminating all repeated and unfeasible solutions (originated if any item remains unstacked after the post-process).


\begin{table}[t]
  \caption{\#sol = number of feasible solutions produced. \\ Runtimes (rt) are in seconds, and for \texttt{QI4WOP} include D-Wave's queue time. For classic method, 30s have been set as maximum runtime. Values are the average after 10 independent runs.}
  \label{tab:first_results}
    \centering
    \resizebox{1.0\columnwidth}{!}{
        \begin{tabular}{l|ll|ll | l|ll|ll}
            \toprule
            \multirow{2}{*}{\bf Instance} & \multicolumn{2}{c|}{\texttt{QI4WOP}} & \multicolumn{2}{c|}{Clas. Ini.} & \multirow{2}{*}{\bf Instance} & \multicolumn{2}{c|}{\texttt{QI4WOP}} & \multicolumn{2}{c}{Clas. Ini.}\\
             & \#sol & rt & \#sol & rt & & \#sol & rt & \#sol & rt\\
            \midrule
            \textbf{L1\_I50\_T2} & 29,9 & 15,7 & 8,5 & 30 & \textbf{L2\_I75\_T3} & 78,9 & 16,7 & 5,8 & 30\\
            \textbf{L3\_I85\_T4} & 96,9 & 17,4 & 4,9 & 30 & \textbf{L3\_I100\_T3} & 117,5 & 17,5 & 4,1 & 30\\
            \textbf{L4\_I124\_T3} & 115,3 & 18,0 & 3,4 & 30 & & & & &\\
            \bottomrule
        \end{tabular}
    }
\end{table}

\section{Preliminary Experimentation}\label{sec:exp}

We conducted a two-phase experimentation to preliminary assess \texttt{QI4WOP}. The objective of the first phase is to compare the efficiency of both the classical \textit{initialization module} (blue steps in Figure \ref{fig:PoC}) and \texttt{QI4WOP} for creating feasible solutions. Five WOP instances have been used\footnote{Data and results available at: \url{http://dx.doi.org/10.17632/h6y87fs5rb.1}}, each named as \texttt{LX\_IY\_TZ}, where \texttt{X} is the number of available locations, \texttt{Y} is the amount of items to store, and \texttt{Z} is the number of item types. We depict in Table \ref{tab:first_results} the outcomes of these first tests, from which we can safely conclude that \texttt{QI4WOP} is able to generate significantly more feasible solutions in remarkably less time. Moreover, while the classic initialization struggles when the size of the instance increases, \texttt{QI4WOP} demonstrates to perform better as long as the complexity of the problem, and thus the solution search space, increases.

The objective of the second phase is to measure the quality of the solutions produced by \texttt{QI4WOP}. For this purpose, we have compared the results obtained by the existing full-classical PoC against the PoC with \texttt{QI4WOP} embedded (by replacing the steps outlined in Figure \ref{fig:PoC}) when solving the biggest instance, \texttt{L4\_I124\_T3}. To draw rigorous conclusions, the execution times of the classical \textit{initialization module} have been significantly increased so that the number of initial solutions generated by both methods is the same. After 25 independent runs, the PoC with \texttt{QI4WOP} embedded has outperformed the full-classical PoC in 48\% of the executions.

As a general conclusion, we can preliminary state that the use of \texttt{QI4WOP} leads to similar performance in terms of results quality, while the savings in terms of computational time are remarkable. This latter aspect can certainly prove to be key in real industrial environments. The results obtained allow us to highlight the work carried out in this research as an example of what quantum computers can offer today, representing the first steps towards obtaining what is known as quantum utility \cite{kim2023evidence}. In order to substantiate these conclusions, further tests will be carried out in the near future.




\bibliographystyle{IEEEtran}
\bibliography{IEEEexample}

\begin{thebibliography}{1}
\providecommand{\url}[1]{#1}
\csname url@samestyle\endcsname
\providecommand{\newblock}{\relax}
\providecommand{\bibinfo}[2]{#2}
\providecommand{\BIBentrySTDinterwordspacing}{\spaceskip=0pt\relax}
\providecommand{\BIBentryALTinterwordstretchfactor}{4}
\providecommand{\BIBentryALTinterwordspacing}{\spaceskip=\fontdimen2\font plus
\BIBentryALTinterwordstretchfactor\fontdimen3\font minus
  \fontdimen4\font\relax}
\providecommand{\BIBforeignlanguage}[2]{{%
\expandafter\ifx\csname l@#1\endcsname\relax
\typeout{** WARNING: IEEEtran.bst: No hyphenation pattern has been}%
\typeout{** loaded for the language `#1'. Using the pattern for}%
\typeout{** the default language instead.}%
\else
\language=\csname l@#1\endcsname
\fi
#2}}
\providecommand{\BIBdecl}{\relax}
\BIBdecl

\bibitem{osaba2024hybrid}
E.~Osaba, E.~Villar-Rodriguez, A.~Gomez-Tejedor, and I.~Oregi, ``Hybrid quantum
  solvers in production: how to succeed in the nisq era?'' \emph{arXiv preprint
  arXiv:2401.10302}, 2024.

\bibitem{leapCQM}
{D-Wave Developers}, ``{Measuring Performance of the Leap Constrained Quadratic
  Model Solver},'' D-Wave Systems Inc., Tech. Rep. 14-1065A-A, 11 2022.

\bibitem{kim2023evidence}
Y.~Kim, A.~Eddins, S.~Anand, K.~X. Wei, E.~Van Den~Berg, S.~Rosenblatt,
  H.~Nayfeh, Y.~Wu, M.~Zaletel, K.~Temme \emph{et~al.}, ``Evidence for the
  utility of quantum computing before fault tolerance,'' \emph{Nature}, vol.
  618, no. 7965, pp. 500--505, 2023.

\end{thebibliography}
\end{document}